\documentclass[pre,aps,twocolumn,amsmath,amssymb]{revtex4-1}
\pdfoutput=1
\usepackage{hyperref}
\usepackage{graphicx}

\usepackage[usenames]{color}
\newcommand{\new}[1]{\textcolor{black}{#1}} 

\def\eqq#1{Eq.~(\ref{#1})}
\def\eq#1{(\ref{#1})}

\def\f#1{Fig.~\ref{#1}}

\def\s#1{Section~\ref{#1}}

\def\c#1{~\cite{#1}}

\def\es{\epsilon_{\rm s}}
\def\en{\epsilon_{\rm n}}

\def\beq{\begin{equation}}
\def\eeq{\end{equation}}
\def\bea{\begin{eqnarray}}
\def\eea{\end{eqnarray}}

\def\kt{k_{\rm B}T}

\def\beq{\begin{equation}}
\def\eeq{\end{equation}}
\def\bea{\begin{eqnarray}}
\def\eea{\end{eqnarray}}

\begin{document}

\title{Minimal physical requirements for crystal growth self-poisoning}
\author{Stephen Whitelam$^1$}
\email[]{swhitelam@lbl.gov}
\author{Yuba Raj Dahal$^2$}
\author{Jeremy D. Schmit$^2$}
\email[]{schmit@phys.ksu.edu}
\affiliation{$^1$Molecular Foundry, Lawrence Berkeley National Laboratory, 1 Cyclotron Road, Berkeley, CA 94720, USA\\
$^2$Department of Physics, Kansas State University, Manhattan, KS, 66506, USA}

\begin{abstract}
Self-poisoning is a kinetic trap that can impair or prevent crystal
growth in a wide variety of physical settings. Here we use dynamic
mean-field theory and computer simulation to argue that poisoning is
ubiquitous because its emergence requires only the notion that a
molecule can bind in two (or more) ways to a crystal; that those
ways are not energetically equivalent; and that the associated
binding events occur with sufficiently unequal probability. If these
conditions are met then the steady-state growth rate is in general a
non-monotonic function of the thermodynamic driving force for
crystal growth, which is the characteristic of poisoning.
Our results also indicate that relatively small changes of system
parameters could be used to induce recovery from poisoning.
\end{abstract}

\maketitle

\section{Introduction}
One of the kinetic traps that can prevent the crystallization of
molecules from solution is the phenomenon of {\em self-posioning},
in which molecules attach to a crystal in a manner not commensurate
with the crystal structure and so impair or prevent crystal
growth\c{ungar2005effect,de2003principles}. This phenomenon has been
seen in computer simulations of hard rods\c{schilling2004self}, and in the
assembly of
polymers\c{higgs1994growth,ungar2000dilution,ungar2005effect} and
proteins\c{asthagiri2000role,schmit2013kinetic}. A signature of self-poisoning
is a growth rate that is a non-monotonic function of the
thermodynamic driving force for crystal growth, with the slowing of
growth as a function of driving force occurring in the
rough-growth-front regime (a distinct effect, growth poisoning at
low driving force, can occur if impurities impair 2D nucleation on
the surface of a 3D
crystal\c{cabrera1958growth,land1999recovery,van1998impurity,sleutel2015mesoscopic}).
Unlike the slow dynamics associated with
nucleation\c{de2003principles,sear2007nucleation}, self-poisoning
cannot be overcome by seeding a solution with a crystal template or
by inducing heterogeneous nucleation.

Here we use dynamic mean-field theory and computer simulation to
argue that poisoning is ubiquitous because its emergence requires no
specific spatial or molecular detail, but only the notion that a
molecule can bind in two (or more) ways to a crystal, optimal and
non-optimal; that the non-optimal way of binding is energetically
less favorable than the optimal way of binding; and that any given
binding event is more likely (by about an order of magnitude) to be
non-optimal than to be optimal. If these conditions are met then the
character of the steady-state growth regime changes qualitatively
with crystal-growth driving force. Just past the solubility limit a
crystal's growth rate increases with thermodynamic driving force
(supercooling or supersaturation). However, the dynamically-generated
crystal also becomes less pure as driving force is increased, i.e.
it incorporates more molecules in the non-optimal configuration. As
a result, the effective driving force for growth of the {\em impure}
crystal can diminish as the driving force for growth of the {\em pure}
crystal increases, and so the impure crystal's growth slows (this
feedback effect is similar to the growth-rate `catastrophe'
described in Ref.\c{van1998impurity}). At even larger driving forces
an impure precipitate of non-optimally-bound molecules grows
rapidly. Self-poisoning of polymer crystallization was studied in
Refs.\c{higgs1994growth,ungar2000dilution,ungar2005effect} using
analytic models and simulations. The present models have a similar
minimal flavor to the models developed in those references, although
our models are not designed to be models of polymer crystallization
specifically, and contain no notion of molecular binding-site
blocking. We show that poisoning can happen even if all molecular
interactions are attractive, and that it results from a nonlinear
dynamical feedback effect that couples crystal quality and crystal
growth rate. Having identified the factors that lead to poisoning,
the present models also suggest that relatively small changes of
system parameters could be used to induce recovery from it.

In \s{sec_mf} we introduce and analyze a mean-field model of the growth of a crystal from molecules able to bind to it in distinct ways. In \s{simulations} we introduce a simulation model of the same type of process, but one that accommodates spatial fluctuations and particle-number fluctuations ignored by the mean-field theory. The behavior of these models is summarized in \s{results}. Both the mean-field model and the simulations show crystal growth rate to be a non-monotonic function of the thermodynamic driving force for growth of the {\em pure} crystal, because the dynamically-generated crystal is in general impure. In some regimes the predictions of the two models differ in their specifics: the mean-field theory assumes a nonequilibrium steady-state of infinite lifetime, and the growth rate associated with this steady-state can vanish. Simulations, which satisfy detailed balance, eventually evolve to thermal equilibrium and so always display a non-zero growth rate. We conclude in \s{conclusions}.

\section{Mean-field theory of growth poisoning}
\label{sec_mf}

The basic physical ingredients of growth poisoning are contained within a model of growth that neglects all spatial detail and accounts only for the ability of particles of distinct type (or, equivalently, distinct conformations of a single particle type) to bind to or unbind from a `structure', which we resolve only in an implicit sense. We consider $K$ types of particle, labeled $i=1,2,\dots,K$ (we will focus shortly on the case of two particle types). We model the structure in a mean-field sense, resolving it only to the extent that we identify the relative abundance $n_i$ of particle type $i$ within the structure, where $\sum_i n_i = 1$ (we assume that sums over variables $i$ and $j$ run over all $K$ particle-type labels). Let us assume that the structure gains particles of type $i$ at rate $p_i C$, where $C$ is a notional concentration and $\sum_i p_i = 1$. Let us assume that particle types unbind from the structure with a rate proportional to their relative abundance within the structure, multiplied by some rate $\lambda$, which can depend on the set of variables $\{n_i\}$. If we write down a master equation for the stochastic process so defined, calculate expectation values of the variables $n_i$, and replace fluctuating quantities by their averages, then we get the following set of mean-field rate equations describing the net rates $\Gamma_i$ at which particles of type $i$ add to the structure:
\beq
\label{rates}
\Gamma_i = p_i C - n_i \lambda(\{n_i\}),
\eeq
where $i=1,2,\dots,K$, and $\sum_i p_i = 1=\sum_i n_i$ as stated previously.  To model a structure of interacting particles we assume a Boltzmann-like rate of unbinding,
\beq
\label{lambdas}
\lambda_i(\{n_i\}) = \exp(\beta \sum_j \epsilon_{ij} n_j),
\eeq
which assumes the interaction energy between particle types $i$ and $j$ to be $\epsilon_{ij}$, and assumes that particles `feel' only the averaged composition $\{n_i\}$ of the structure.

We define the growth rate of the structure as
\beq
\label{growth_rate}
V\equiv\sum_i \Gamma_i.
\eeq
In equilibrium the structure neither grows nor shrinks, and we have
\beq
\label{equilib}
\Gamma_i=0
\eeq for each $i=1,2,\dots,K$. We shall also assume the existence of a steady-state growth regime in which $V \geq 0$ but the composition of the structure does not change with time; in this regime we have
\beq
\label{steady}
n_i=\frac{\Gamma_i}{\sum_j \Gamma_j},
\eeq
i.e. the relative abundance of each particle type is proportional to the relative rate at which it is added to the structure.

At this point the set of equations \eq{rates} -- \eq{steady} describes a generic model of growth via the binding and unbinding of particles of multiple types. The model is mean-field in both a spatial sense --  no spatial degrees of freedom exist, and particle-structure interactions depend on the composition of the structure as a whole -- and in the sense of ignoring fluctuations of particle number: the model resolves only net rates of growth. We now specialize the model to the case of crystal growth in the presence of impurities; different choices of parameters can be used to model other scenarios\c{Whitelam2014a,Sue2015,mannige2015predicting}.

We shall consider two particle types, and so set $K=2$. We will call particle types 1 and 2 `B' for `blue' and `R' for `red', respectively, for descriptive purposes (in Section~\ref{results} simulation configurations will be color-coded accordingly). We call the relative abundance of blue particles in the structure $n_1 \equiv n$, and  so the relative abundance of red particles in the structure is $n_2=1-n$. We consider blue particles to represent the (unique) crystallographic orientation and conformation of a particular molecule, and red particles to represent the ensemble of non-crystallographic orientations and conformations of the same molecule. Alternatively, one could consider red particles to be an impurity species present in the same solution as the blue particles that we want to crystallize. We assume that an isolated particle is blue with probability $p$ and red with probability $1-p$, and so we choose $p_1 = p$ and so $p_2 = 1-p$ for the basic rates of particle addition in \eq{rates}. We will assume that the blue-blue crystallographic or `specific' interaction in \eqq{lambdas} is $\epsilon_{\rm BB}=-\es \kt$. We will assume that interactions between blue and red ($\epsilon_{\rm RB}$) or red and red ($\epsilon_{\rm RR}$) are `nonspecific', and equal to $-\en \kt$. With these choices \eq{rates} reads
 \bea
 \label{rateb}
\Gamma_{\rm B} &=& pC -n \alpha^n {\rm e}^{- \en}, \\
\label{rater} \Gamma_{\rm R} &=& (1-p) C -(1-n) {\rm e}^{- \en},
 \eea
 where $\alpha \equiv {\rm e}^{-\Delta}$ and $\Delta  \equiv \es -\en$. This model describes the growth of a structure whose character is defined by its `color', $n$; for $n\approx 1$ the structure is almost blue, and we shall refer to this structure as the `crystal'. For $n$ small we have a mostly red structure, and we refer to this as the `precipitate'. Intermediate values of $n$ describe a structure that we shall refer to as an `impure' crystal.

It is convenient to work with a set of rescaled rates and concentrations
\beq
 \label{rescale}
 \left(c, \gamma_{\rm R}, \gamma_{\rm B} \right) \equiv \left(C,\Gamma_{\rm R}, \Gamma_{\rm B} \right) {\rm e}^{\en},
 \eeq
  in terms of which Equations~\eq{rateb} and~\eq{rater} read
 \bea
 \label{ratebreduced}
 \gamma_{\rm B} &=& p c -n \alpha^n.\\
 \label{raterreduced}
\gamma_{\rm R} &=& (1-p)c -(1-n).
 \eea
The rescaling defined by~\eqq{rescale} makes an important physical
point: the timescale for crystal growth is measured most naturally
in terms of the basic timescale ${\rm e}^{\en}$ for the unbinding of
impurity (red) particles. Thus, for fixed energy scale $\en \kt$,
lowering temperature serves to increase this basic timescale,
indicating that cooling is not necessarily a viable strategy for
speeding crystal growth.

Equation \eq{steady}, which reduces to
\beq
\frac{n}{1-n} = \frac{\gamma_{\rm B}}{\gamma_{\rm R}},
\eeq
is the assumption that there exists a steady-state dynamic regime in which the relative abundance of red and blue particles in the growing structure is equal to the ratio of their rates of growth. Inserting into this condition Equations \eq{ratebreduced} and \eq{raterreduced} gives the self-consistent relation
\beq
 \label{noneqcomposition}
\frac{n}{1-n} = \frac{p c - n \alpha^n}{(1-p)c -(1-n)}.
\eeq
One can solve this equation graphically for solid composition $n$, as a function of the parameters $\es, \en, c$, and $p$. To determine the growth rate of the solid one inserts the value of $n$ so calculated into Equations \eq{ratebreduced} and \eq{raterreduced}, and adds them:
 \beq
 \label{vee}
 v=\gamma_{\rm R}+\gamma_{\rm B}.
 \eeq
The physical growth rate is then $V = v {\rm e}^{-\en}$, obtained by undoing the rescaling~\eq{rescale}.
\begin{figure}
\includegraphics[width=\linewidth]{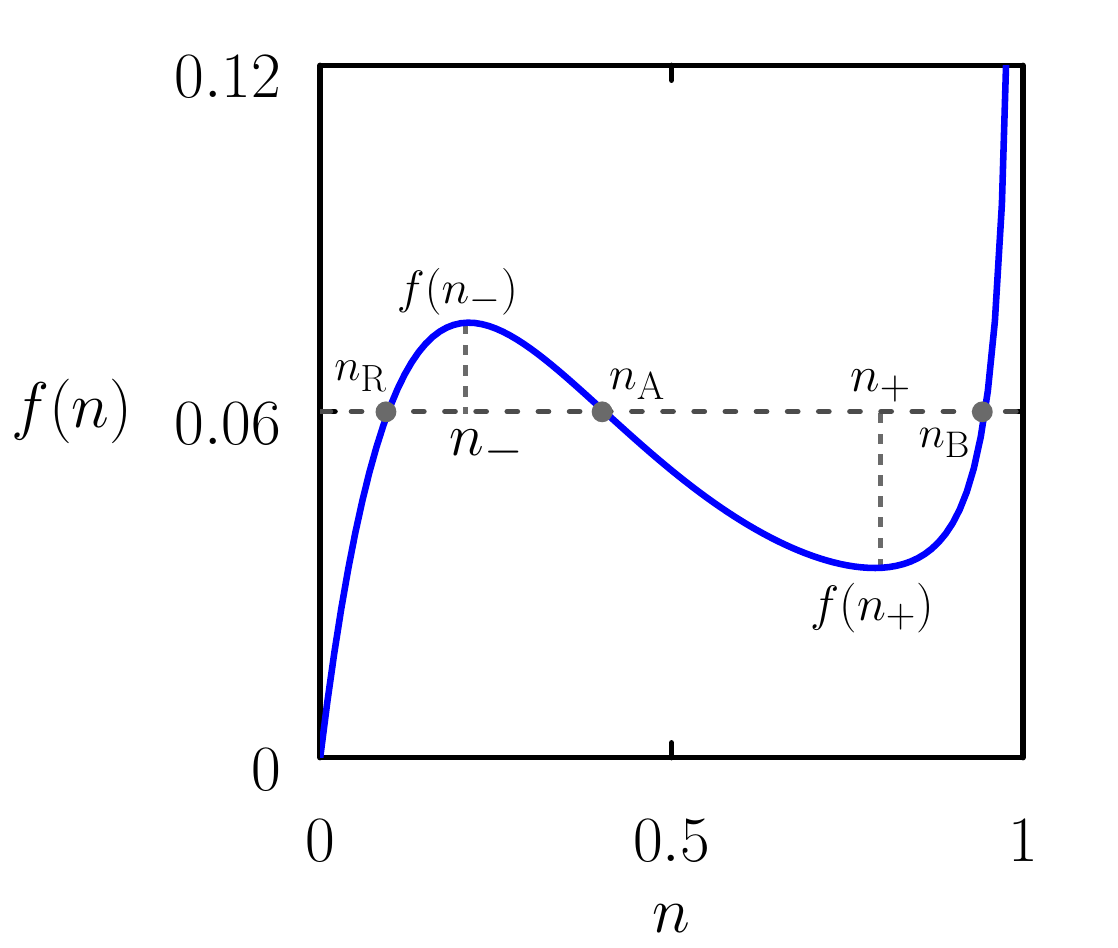}
\caption{\label{fig_graphical} Graphical construction used to determine the phase diagram of the mean-field model of growth poisoning (see \f{fig_mf}(a)). The solutions of \eqq{eqstate} give the solid compositions at which the growth rate vanishes. The horizontal dotted line shows a value of $p/(1-p)$ for which three such solutions exist; the associated values $n_{\rm B}$, $n_{\rm A}$ and $n_{\rm R}$ lie on the `solubility', `arrest', and `precipitation' lines shown in \f{fig_mf}(a).}
\end{figure}

To gain insight into the behavior of the model is it useful to solve \eqq{noneqcomposition} for $c$,
\beq
\label{cparam}
c=\frac{n(1-n)}{n-p} \left(1-\alpha^n\right),
\eeq
and to use this expression to eliminate $c$ from \eq{vee}, giving
\beq
\label{vparam}
v=\frac{1}{n-p} \left[ (1-n)p-n(1-p) \alpha^n\right].
\eeq
Equations \eq{cparam} and \eq{vparam} can be regarded as parametric equations for the concentration $c$ at which one observes a particular growth rate $v$ of a solid of composition $n$~\footnote{\new{Note that the limit $n \to p$ can be obtained either when blue and red are energetically equivalent, i.e. when $\alpha \to 1$, or in the `solid solution' limit of rapid deposition. In the first case \eqq{noneqcomposition} shows that $n=p$ for any concentration: blue is added to the structure in proportion $p$, and because red and blue are energetically equivalent there exists no mechanism, at any growth rate, to change that proportion. In this case \eq{vparam} is ill-defined, but adding \eq{ratebreduced} and \eq{raterreduced} yields the growth rate $v=c-1$. The solid solution limit is obtained when $c \to \infty$, in which case $v \to c \to \infty$. In this case the expressions \eq{cparam} and \eq{vparam} are singular, and these singularities are physically appropriate.}}. \new{Note that $v$ can be negative for certain parameter combinations, indicating a breakdown of the assumption of a steady-state growth regime.} The basic phenomenology revealed by Equations \eq{cparam} and \eq{vparam} is that altering concentration $c$ results in a change of composition $n$ of the growing structure, and that changes of both $c$ and $n$ affect the rate of growth $v$.

\begin{figure*}
\includegraphics[width=\linewidth]{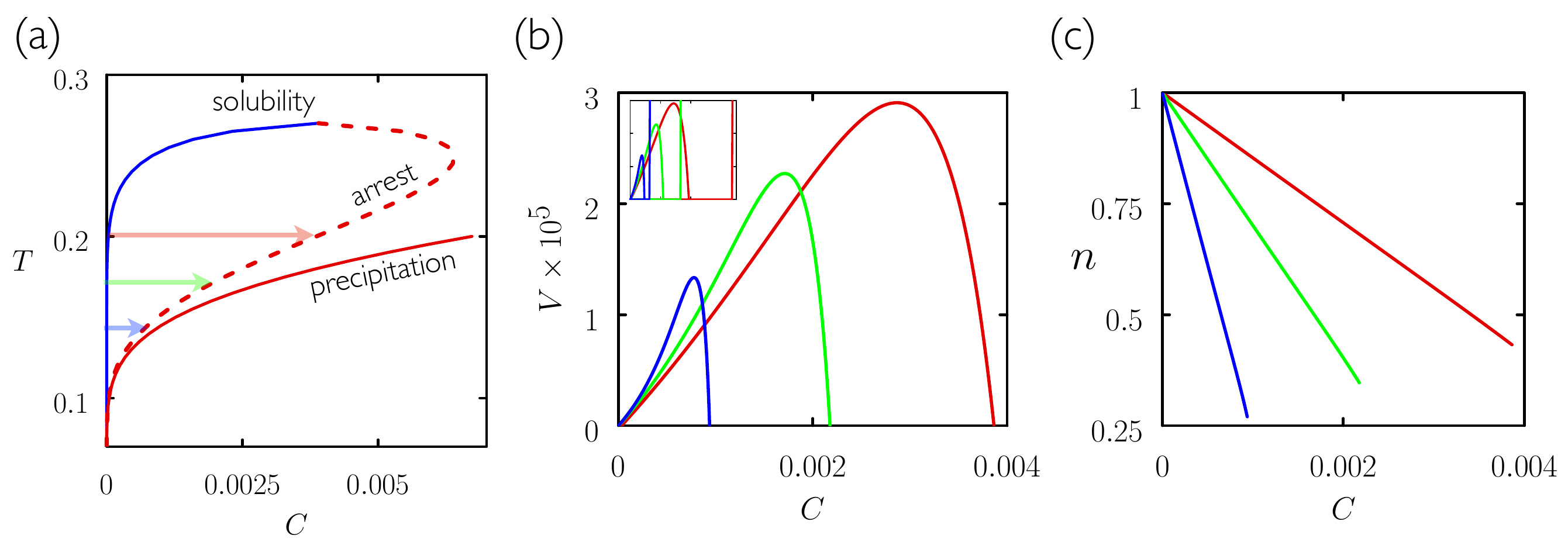}
\caption{\label{fig_mf} Dynamic mean-field theory predicts that crystal growth rate is a non-monotonic function of concentration. (a) Mean-field phase diagram in the temperature $(T)$-concentration $(C)$ plane derived from Equations \eq{eqstate} and \eq{eqconc}. The line marked `solubility' shows the concentration at which the crystal (the `blue' solid) neither grows nor shrinks; the line marked `precipitation' is the same thing for the impure (`red') precipitate. The line marked `arrest' shows where the growth rate of the (impure) crystal goes to zero. (b) Growth rate $V$ and (c) crystal quality $n$ as a function of concentration at the three temperatures indicated in the left panel (line colors correspond to arrow colors), obtained from Equations \eq{noneqcomposition} and \eq{vee}. At the solubility line the crystal does not grow; upon supersaturation it grows with finite speed $V$ and becomes less pure. Consequently, its growth rate begins to decline for sufficiently large $C$, going to zero at the arrest line. Beyond the precipitation line the precipitate grows rapidly (see inset in (b), drawn as in the main panel but with $C$ extended to just beyond the precipitation line). Parameters: $p=10^{-2}$; $\Delta/T \equiv (\es-\en)/T = 2/T$; $ \en/T = 1/T$.}
\end{figure*}

In certain parameter regimes $v$ can become a non-monotonic
function of $c$, which is crystal growth poisoning. This potential
can be seen from \eq{vparam}; setting $\partial v/\partial n = 0$
yields~\footnote{\new{The condition $\partial v/\partial n = 0$ is also satisfied when $p=1$, the limit in which only blue particles are added to the structure. In this case $v=c-1$, independent of $n$}.}
\beq
  p = \frac{n^2 \Delta}{{\rm e}^{n \Delta}-1+n \Delta}.
\eeq
 The right-hand side of this equation is a non-monotonic function of $n$, and takes
its maximum value when $n=2/\Delta$. Thus for $\Delta>2$ this
equation has two solutions (equivalent to turning points of $v(n)$)
provided that $\Delta(e^\Delta - 1 +\Delta)^{-1}<p<4
\Delta^{-1}(1+{\rm e}^2)^{-1}$. These two solutions underpin the
behavior shown in \f{fig_mf}: increasing concentration first causes
the structure to grow more rapidly (because we increase the driving
force for crystal growth), and then more slowly (as poisoning
happens), and then more quickly again (as the structure grows in an
`impure' way). For $\Delta =4$ (see below) poisoning happens if
$p<(1+{\rm e}^2)^{-1} \approx 0.12$. That is, for poisoning to happen the impure (red) species
must be at least about 10 times more abundant in solution than the
crystal-forming (blue) species.

Of particular interest are the locations in phase space where the growth rate vanishes. These locations can be identified by setting the right-hand side of \eq{vparam} to zero (or equivalently setting $\gamma_{\rm B}=\gamma_{\rm R}=0$ in Equations \eq{ratebreduced} and \eq{raterreduced}), giving
\beq
\label{eqstate}
\frac{p}{1-p} = f(n)
\eeq
where
\beq
\label{eq_func}
f(n) = \frac{n}{1-n} \alpha^n.
\eeq
Recall that $\alpha \equiv {\rm e}^{-\Delta}$ and $\Delta \equiv \es - \en$. Inspection of the properties of $f(n)$ reveals the conditions under which growth arrest can occur. To this end it is convenient to calculate the stationary points $n_\pm$ of $f(n)$, which are
\beq
n_\pm = \frac{1}{2} \left( 1 \pm \sqrt{1-4 \Delta^{-1}}\right).
\eeq
Two stationary points exist for $\Delta >4$, where the function $f(n)$ has the behavior shown in \f{fig_graphical}. Arrest can happen if the horizontal line $p/(1-p)$ lies between the values $f(n_-)$ and $f(n_+)$, i.e. if
\beq
\frac{f(n_-)}{1+f(n_-)} > p > \frac{f(n_+)}{1+f(n_+)},
\eeq
where
\beq
f(n_\pm) = \frac{1\pm\chi}{1\mp \chi} \exp\left\{ -\frac{\Delta}{2} \left( 1\pm \chi\right)\right\},
\eeq
with $\chi \equiv \sqrt{1-4 \Delta^{-1}}$. In this case there are three solutions $n_\theta$ to \eqq{eqstate}. We shall call these solutions $n_{\rm B}$, $n_{\rm A}$, and $n_{\rm R}$. From \eqq{raterreduced} the associated concentrations $c_\theta$ are
\beq
\label{eqconc}
c_\theta= \frac{1-n_\theta}{1-p},
\eeq
 where $\theta = $ R, B, or A. The solution corresponding to
the largest value of $n$ we call $n_{\rm B}$ (B for blue). The
associated concentration $c_{\rm B}$ is that at which the
mostly-blue solid or `crystal' is in equilibrium, and we shall call
the locus of such values, calculated for different parameter
combinations, the `solubility line'. The solution corresponding to
the smallest value of $n$ we call $n_{\rm R}$ (R for red). The
associated concentration is that at which the mostly-red
`precipitate' is in equilibrium, and this lies on what we will call
the `precipitation line'. The remaining solution we call $n_{\rm A}$
(A for arrest); it yields the concentration at which the impure
crystal ceases to grow, and it lies on the `arrest line'. 

Arrest therefore occurs when $\Delta$ is large
enough that the (blue) crystal is stable thermodynamically {\em and} $p$ is small enough that the crystal's emergence is kinetically hindered. If $p$ is large enough, i.e. if $p/(1-p)$ is greater than
$f(n_-)$, then the crystal's emergence is not kinetically hindered and growth arrest does not occur. Conversely, if $p$ is too small, i.e. if $p/(1-p)$ is less than $f(n_+)$, then $\Delta$ is too small to render the crystal thermodynamically stable.

We can use this set of equations to determine the behavior of our mean-field model of crystal growth, and we describe this behavior in \s{results}. There we revert to `physical' growth rates $V$ and concentrations $C$; these are related to their rescaled counterparts $v$ and $c$ via \eqq{rescale}.

\section{Computer simulations of two-component growth}
\label{simulations}
\begin{figure*}
\includegraphics[width=0.8\linewidth]{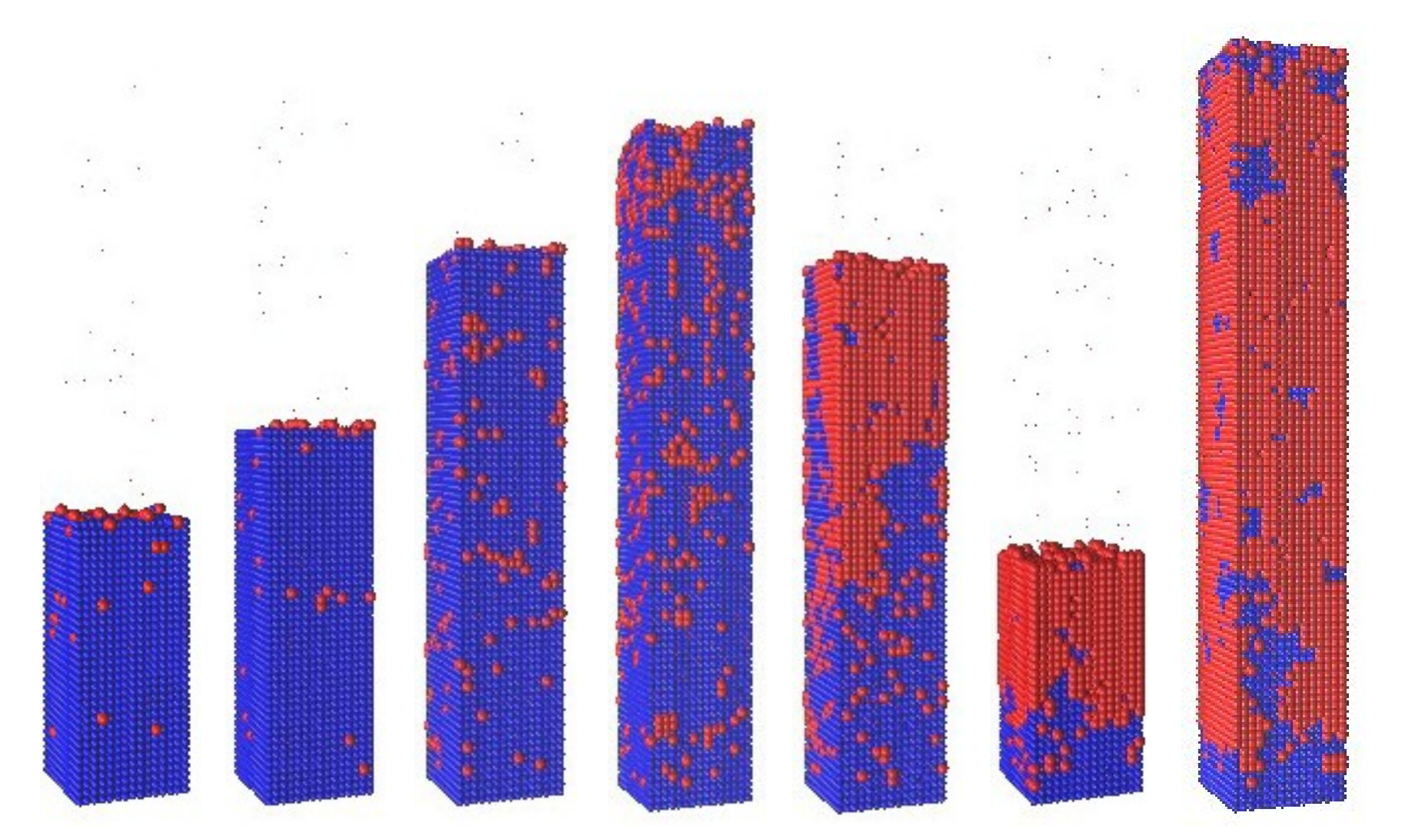}
\caption{\label{fig_sim_snapshot} Simulation snapshots taken after
fixed long times ($5\times 10^9$ MC sweeps) for a range of
concentrations $c$ (increasing from left to right) bear out the key
prediction of mean-field theory: growth rate is a non-monotonic
function of the driving force for crystal growth. Growth rate first
increases and then decreases with concentration, because the growing
structure becomes less pure (more red). The right-hand snapshot lies
beyond the precipitation line, where the impure solid grows rapidly.
Parameters: $p=10^{-2}$; $\es=3.5$; $\en=1.4$. From left to right,
values of $c$ are $0.008, 0.0083, 0.009875, 0.0119, 0.014225,
0.0149, 0.01512$.}
\end{figure*}

We carried out lattice Monte Carlo simulations of two-component
growth, similar to those done in
Refs.\c{Whitelam2014a,Hedges2014,Sue2015,mannige2015predicting}.
Simulations, which satisfy detailed balance with respect to a
particular lattice energy function, accommodate spatial degrees of
freedom and particle-number fluctuations omitted by the mean-field
theory. Simulations therefore provide an assessment of which physics
is captured by the mean-field theory and which it omits.

Simulation boxes consisted of a 3D cubic lattice of $15\times15\times100$ sites. Sites can be vacant (white), or occupied by a blue particle or a red particle. Periodic boundary conditions were applied along the two short directions. At each time step a site was chosen at random. If the chosen site was white then we proposed with probability $p$ to make it blue, and with probability $1-p$ to make it red. If the chosen site was red or blue then we proposed to make it white. No red-blue interchange was allowed. To model the slow dynamics in the interior of an aggregate we allowed no changes of state of any lattice site that had 6 colored nearest neighbors.

These proposals we accepted with the following probabilities:
\begin{eqnarray}
\textrm{R} \to \textrm{W} &:& \min\left(1,(1 - p){\rm e}^{-\beta \Delta E}\right); \nonumber \\
\textrm{W} \to \textrm{R} &:& \min\left(1,(1 - p)^{-1}{\rm e}^{-\beta\Delta E}\right); \nonumber \\
\textrm{B} \to \textrm{W} &:& \min\left(1,p \,{\rm e}^{-\beta\Delta E} \right); \nonumber \\
\textrm{W} \to \textrm{B} &:& \min\left(1,p^{-1} {\rm e}^{-\beta \Delta E}  \right), \nonumber
\end{eqnarray}
where $\Delta E$ is the energy change resulting from the proposed move. This change was calculated from the lattice energy function
\begin{eqnarray}
E =  \sum_{<i,j>} \epsilon_{C(i)C(j)} +  \sum_{i} \mu_{C(i)} \label{E}.
\end{eqnarray}
The first sum runs over all distinct nearest-neighbor interactions and the second sum runs over all sites. The index $C(i)$ describes the color of site $i$, and is W (white), B (blue) or R (red); $\epsilon_{C(i)C(j)}$ is the interaction energy between colors $C(i)$ and $C(j)$; and the chemical potential $\mu_{C(i)}$ is $\mu \kt$, $-\kt \ln p$  and $-\kt \ln(1-p)$ for W, B and R, respectively (note that positive $\mu$ favors particles over vacancies). In keeping with the choices made in \s{sec_mf} we take
\bea
\epsilon_{\rm BB}= -\es \kt;\\
\epsilon_{\rm BR} = \epsilon_{\rm RB}=\epsilon_{\rm RR}= -\en \kt.
\eea
In the absence of pairwise energetic interactions the likelihood that a given site will be white, blue or red is respectively $1/(1+{\rm e}^{\mu})$, $p/(1+{\rm e}^{-\mu})$, and $(1-p)/(1+{\rm e}^{-\mu})$.

Simulations were begun with three complete layers of blue particles at one end of the box to eliminate the need for spontaneous nucleation. For fixed values of energetic parameters we measured the composition $n$ (the fraction of colored blocks that are blue) and growth rate of the structure produced at different values of the parameter $c \equiv {\rm e}^{\mu}$ (which for small $c$ is approximately equal to the likelihood than an isolated site will in equilibrium be colored rather than white).
\begin{figure*}
\includegraphics[width=0.8\linewidth]{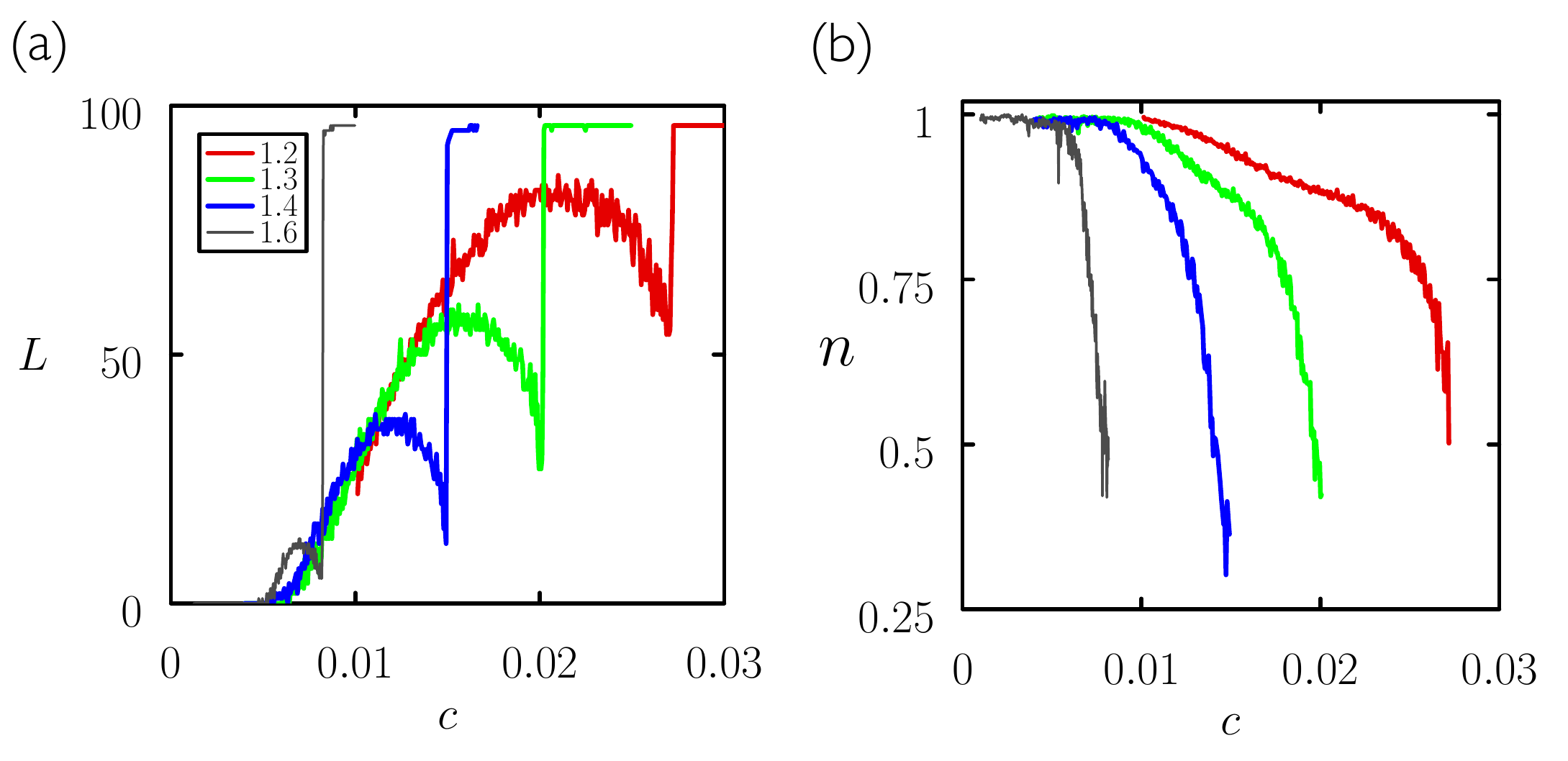}
\caption{\label{fig_sim_plot} Simulations show a non-monotonic growth rate and decline of crystal quality as the driving force for crystal growth is increased. Number of deposited layers $L$ (a) and crystal quality $n$ (b) after $5\times 10^9$ MC sweeps as a function of concentration $c$, for simulations run at various values of the nonspecific interaction parameter $\en$ (plot legends show values of $\en$). The spike in growth rate at large $c$ signals the passing of the precipitation line. Other parameters: $p=10^{-2}$; $\es=3.5$.}
\end{figure*}

\section{Results}
\label{results}

\f{fig_mf}(a) shows the phase diagram of our mean-field model of crystal growth. The `solubility' and `precipitation' lines indicate where the crystal and precipitate are in equilibrium; the `arrest' line shows where the growth rate of the impure crystal vanishes. The structure of this diagram is similar to that of certain experimental systems -- see e.g. Refs.\c{Asherie2004,Luft2011} or Fig. 14 of Ref.\c{ungar2005effect} -- showing that the mean-field theory, although simple, can capture important features of real systems. Upon moving left to right across this diagram we observe the behavior shown in panels (b) and (c) of the figure. Growth rate $V$ first increases with concentration $C$, because the thermodynamic driving force for crystal growth increases. But at some point $V$ begins to decrease, i.e. poisoning occurs. This is so because the composition of the growing solid changes with concentration -- it becomes less pure -- and so the thermodynamic driving force for its growth decreases, even through the thermodynamic driving force for the growth of the {\em pure} crystal increases with $C$. As we pass the precipitation line the growth rate $V$ becomes large and positive (inset to panel (b)). This behavior is similar to that seen in e.g. Fig. 2 of Ref.\c{ungar2000dilution}.

The mean-field theory is simple in nature but furnishes
non-trivial predictions. Key aspects of these predictions are borne
out by our simulations, which resolve spatial detail and
particle-number fluctuations omitted by the theory (we found similar
theory-simulation correspondence in a different regime of parameter
space\c{Whitelam2014a}). In \f{fig_sim_snapshot} we show simulation
snapshots, taken after fixed long times, for a range of values of
concentration $c$. One can infer from this picture that growth rate
is a non-monotonic function of concentration. In all cases the
equilibrium structure is a box mostly filled with blue particles. At
small concentrations we see the growth of a structure similar to the
equilibrium one. Poisoning occurs because the grown structure
becomes less pure (more red) as $c$ increases, and so the effective
driving force for its growth decreases even though the driving force
for the growth of the pure crystal increases. At large
concentrations we pass the precipitation line and the impure (red)
solid grows rapidly.

In \f{fig_sim_plot} we show the number of layers $L$ deposited after
fixed long simulation times for various concentrations $c$ (we
consider a layer to have been added if more than half the sites in a
given slice across the long box direction are are occupied by red or
blue particles). The general trend seen in simulations is similar to
that seen in the mean-field theory (panels (b) and (c) of
\f{fig_mf}). At concentrations just above the blue solubility limit
the structure's growth rate increases approximately linearly with
concentration. At higher concentrations the growth rate reaches a
maximum and then drops sharply, because structure quality (and so
the effective driving force for its growth) declines with
concentration. One difference between mean-field theory and
simulations is that in the latter the growth rate in the poisoning
regime does not go to zero. This is so because simulations satisfy
detailed balance, and must eventually evolve to equilibrium.
Fluctuations (mediated within the bulk of the structure by
vacancies) allow the composition of an arrested structure to evolve
slowly toward equilibrium, and thereby to extend slowly. Thus the
steady-state dynamic regime that has infinite lifetime with the
mean-field theory has only finite lifetime within our simulations
(because these eventually must reach equilibrium). Slow evolution of
this nature is shown in \f{coarsening}.

\begin{figure*}[]
\includegraphics[width=0.7\linewidth]{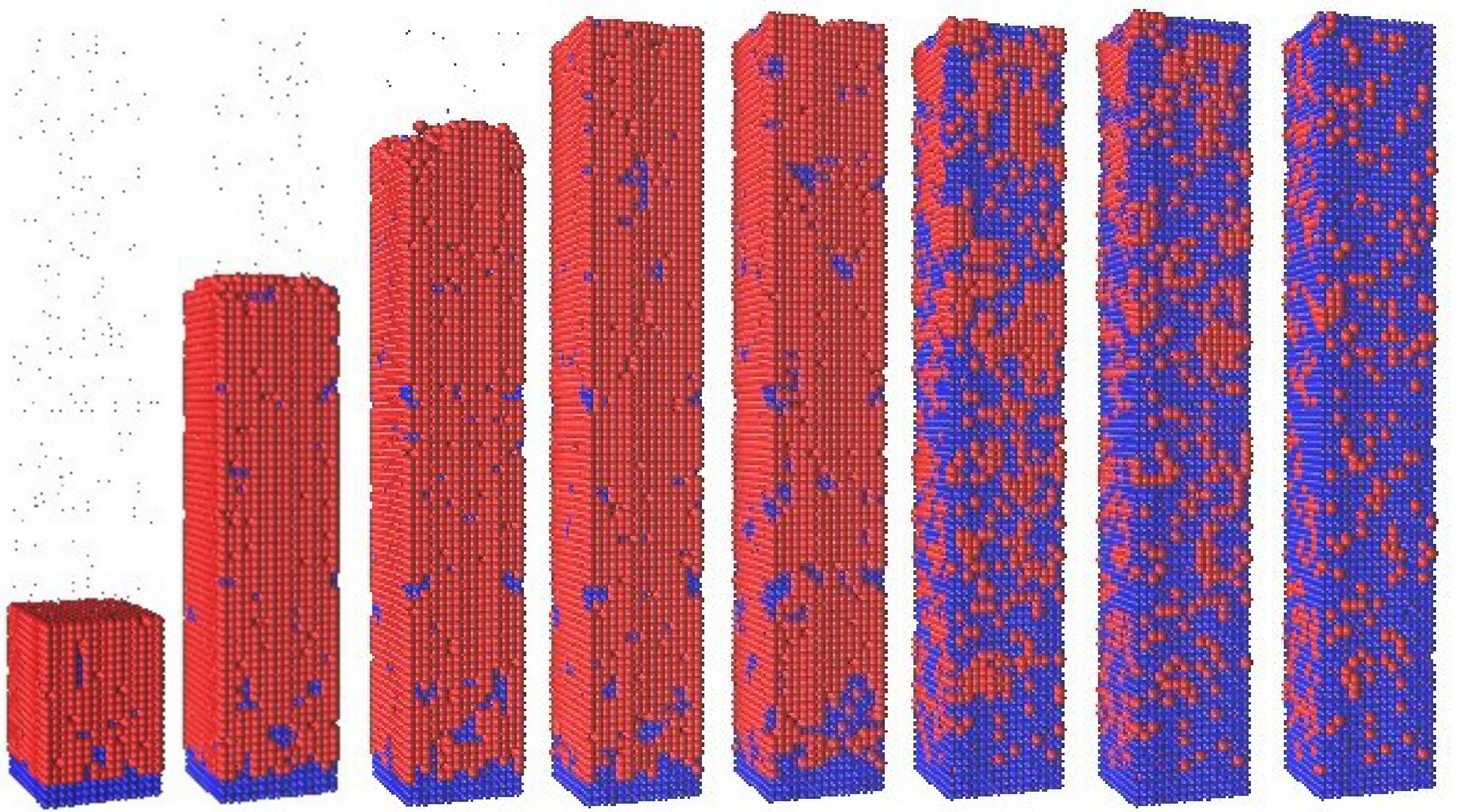}
\caption{Simulations satisfy detailed balance and so eventually
evolve to equilibrium. Here we show a time-ordered series of
snapshots from a simulation done within the precipitation
regime. Fluctuations allow the eventual emergence of the
thermodynamically stable crystal structure. Parameters: $p=10^{-2}$;
$\es=3.5$; $\en=1.2$; $c=0.0274$.} \label{coarsening}
\end{figure*}

\section{Conclusions}
\label{conclusions}

We have used mean-field theory and computer simulation to show that
crystal growth self-poisoning requires no particular spatial or
molecular detail, as long as a small handful of physical ingredients
are realized. These ingredients are: the notion that a molecule can
bind in two (or more) ways to a crystal; that those ways are not
energetically equivalent; and that they are realized with
sufficiently unequal probability. If these conditions are met then
the steady-state growth rate of a structure is, in general, a
non-monotonic function of the thermodynamic driving force for
crystal growth. Self-poisoning is seen in a wide variety of physical
systems\c{schilling2004self,ungar2005effect,asthagiri2000role},
because, we suggest, many  molecular systems display the three
physical ingredients we have identified as being sufficient
conditions for poisoning. Protein crystallization, for instance, is
notoriously difficult, and rational guidance for it is much
needed\c{ten1997enhancement,george1994predicting,shim2007using,haxton2012design,schmit2012growth,fusco2015soft}.
\new{The present model suggests that proteins are prime candidates for
self-poisoning because they have smaller effective values of the $p$
parameter (which controls the relative rates of binding of optimal and
non-optimal contacts) than do relatively rigid small molecules: proteins are anisotropic, conformationally
flexible objects whose non-crystallographic modes of binding
outnumber their crystallographic mode of binding by a factor of
order $10^4$ or
$10^5$\c{Kierzek1997, asthagiri2000role,schmit2012growth}}. Many
protein crystallization trials result in clear solutions without any
obvious indication of why crystals failed to appear
\cite{Luft2011a}, and in some of these cases self-poisoning might be
happening. \new{In general terms decreasing $p$ leaves a system vulnerable to poisoning because a) the rate of attachment of non-crystallographic conformations increases, and b) to ensure thermodynamic stability of the crystal one must increase the basic binding energy scale, in which case the basic timescale for growth increases.}

There also exists a possible connection between the present work and the recent observation of protein clusters that appear in weakly-saturated solution and do not grow or shrink\c{pan2010origin}. Other authors have proposed\c{pan2010origin} and formulated\c{lutsko2015mechanism} models that explain the long-lived nature of such clusters via the slow interconversion of oligomeric and monomeric protein: in these models there exists a thermodynamic driving force to grow clusters of oligomers, but the growth of such clusters is hindered by the existence of monomeric protein. If we reinterpret the present model to regard the `red' species as monomeric protein and the `blue' species as oligomeric protein, then we obtain a possible connection to the mechanism described in Refs.\c{pan2010origin, lutsko2015mechanism}. From e.g. \f{fig_mf}(a) we see that we can be in a region of phase space that is undersaturated with respect to monomeric (red) protein but supersaturated with respect to oligomeric (blue) protein (i.e. the thermodynamic ground state is a condensed structure built from oligomeric protein). There then exists a thermodynamic driving force to grow structures made of oligomeric protein, but the emergence of such structures is rendered slow by kinetic trapping (caused by the fact that monomeric protein is more abundant in isolation than is oligomeric protein). According to this interpretation the `stable' protein clusters are kinetically trapped, and on long enough timescales would grow. However, we stress that this connection is tentative. 

Having identified factors that lead to poisoning, the present models also suggest that relatively small changes of system parameters could be used to avoid it. For instance, \f{fig_mf} and \f{fig_sim_plot} show that, given a set of molecular characteristics, small changes of concentration or temperature can take one from a poisoned regime to one in which crystal growth rate is relatively rapid. Recovery from poisoning could also be effected if one has some way of altering molecular characteristics, such as the value of the non-optimal binding energy scale; see \f{fig_sim_plot} and \f{fig_mf_supp}.

\begin{figure}[]
\includegraphics[width=\linewidth]{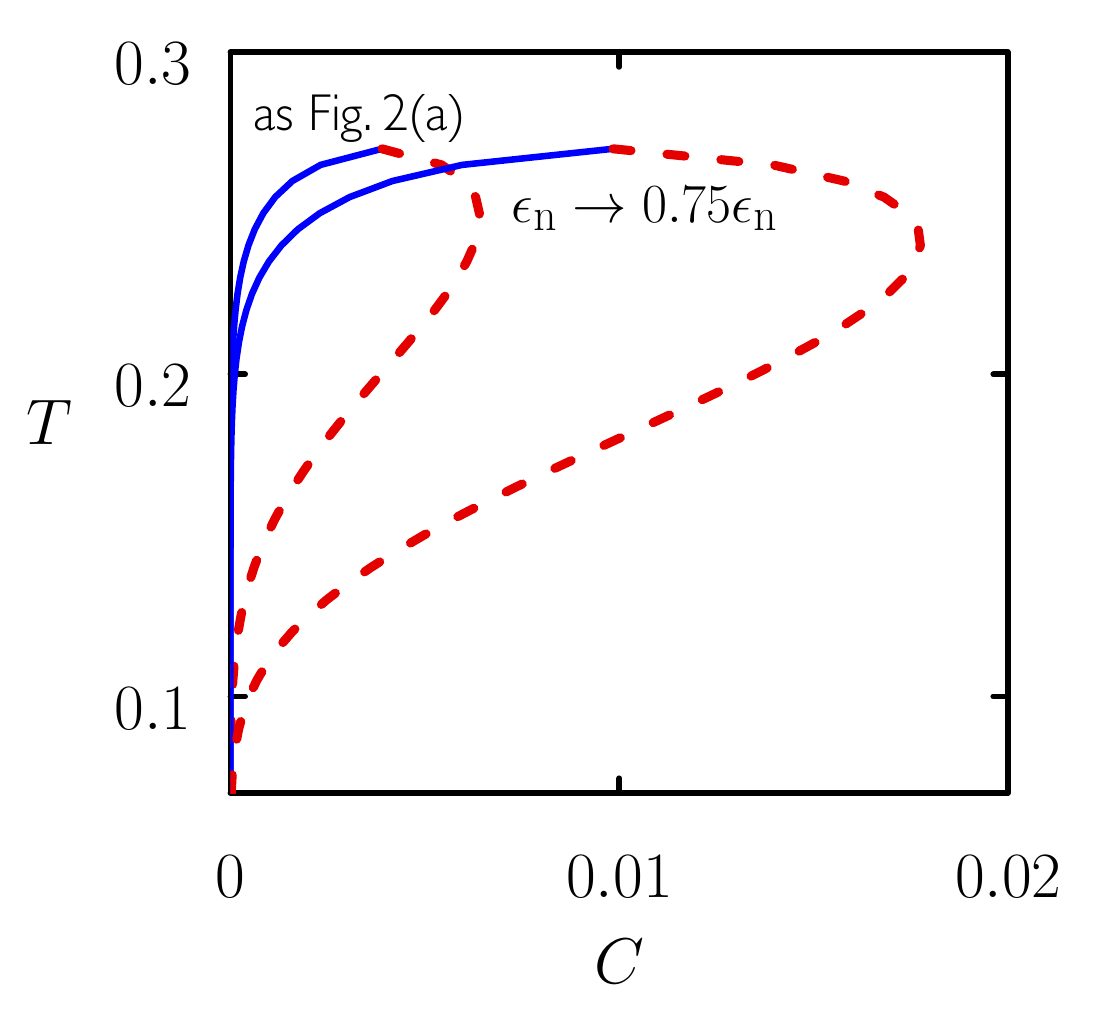}
\caption{Solubility and arrest line calculated from mean-field theory as in \f{fig_mf} (with no precipitation line drawn), with a second solution, the larger loop to the right, drawn for the case of diminished nonspecific binding energy $\en \to 3\en/4$ (with $\Delta$ unchanged). This change greatly enlarges the region of phase space in which crystal growth can happen.}
\label{fig_mf_supp}
\end{figure}

\acknowledgements

This work was done as part of a User project at the Molecular Foundry at Lawrence Berkeley National Laboratory, supported by the Office of Science, Office of Basic Energy Sciences, of the U.S. Department of Energy under Contract No. DE-AC02--05CH11231. JDS would like to acknowledge support from NIH Grant R01GM107487. Computer facilities were provided by the Beocat Research Cluster at Kansas State University, which is funded in part by NSF grants CNS-1006860, EPS-1006860,  and EPS-0919443.


%

\end{document}